# One-stop Automated Diagnostic System for Carpal Tunnel Syndrome in Ultrasound Images Using Deep Learning

*Jiayu Peng[1,2#], Jiajun Zeng[3,4,5#], Manlin Lai[6#], Ruobing Huang[3,4,5], Dong Ni[3,4,5], Zhenzhou Li[1,2]\**

[1]Department of Ultrasound, The Second People's Hospital of Shenzhen, The First Affiliated Hospital of Shenzhen University, Shenzhen, China

[2]Shenzhen University Health Science Center, Shenzhen, 518000, P.R. China

[3]National-Regional Key Technology Engineering Laboratory for Medical Ultrasound, School of Biomedical Engineering, Health Science Center, Shenzhen University, Shenzhen, China

[4]Medical Ultrasound Image Computing (MUSIC) Lab, Shenzhen University, China

[5]Marshall Laboratory of Biomedical Engineering, Shenzhen University, China

[6]Department of Medical Imaging (DMI) - Ultrasound Division, The University of Hong Kong-Shenzhen Hospital, Shenzhen, China

\* Correspondence: Zhenzhou Li, Department of Ultrasound, The Second People's Hospital of Shenzhen, The First Affiliated Hospital of Shenzhen University, No. 3002, Sungang West Road, Futian District, Shenzhen, China, 518061, E-mail: lizhenzhou2004@126.com

# Jiayu Peng, Jiajun Zeng and Manlin Lai contributed equally to this manuscript.

**Abstract**

Objective: Ultrasound (US) examination has unique advantages in diagnosing carpal tunnel syndrome (CTS) while identifying the median nerve (MN) and diagnosing CTS depends heavily on the expertise of examiners. To alleviate this problem, we aimed to develop a one-stop automated CTS diagnosis system (OSA-CTSD) and evaluate its effectiveness as a computer-aided diagnostic tool.

Methods: We combined real-time MN delineation, accurate biometric measurements, and explainable CTS diagnosis into a unified framework, called OSA-CTSD. We collected a total of 32,301 static images from US videos of 90 normal wrists and 40 CTS wrists for evaluation using a simplified scanning protocol.

Results: The proposed model showed better segmentation and measurement performance than competing methods, reporting that HD95 score of 7.21px, ASSD score of 2.64px, Dice score of 85.78%, and IoU score of 76.00%, respectively. In the reader study, it demonstrated comparable performance with the average performance of the experienced in classifying the CTS, while outperformed that of the inexperienced radiologists in terms of classification metrics (e.g., accuracy score of 3.59% higher and F1 score of 5.85% higher).

Conclusion: The OSA-CTSD demonstrated promising diagnostic performance with the advantages of real-time, automation, and clinical interpretability. The application of such a tool can not only reduce reliance on the expertise of examiners, but also can help to promote the future standardization of the CTS diagnosis process, benefiting both patients and radiologists.

**Introduction**

Carpal tunnel syndrome (CTS) is the most frequently encountered type of peripheral compression neuropathy characterized by median nerve (MN) entrapment at the wrist(1). The carpal tunnel is bounded by the transverse carpal ligament on the volar side and eight carpal bones on the dorsal side. Nine flexor digital tendons and the MN pass through the carpal tunnel in the wrist. In CTS, nerve compression causes local circulatory disorders, damage to the blood-nerve barrier, increased pressure of the nerve endoneurial fluid, edema and thickening of blood vessel walls, non-inflammatory synovial fibrosis and vascular proliferation, fibrosis, and thinning of the nerve myelin sheath, resulting in a series of clinical symptoms such as pain, sensory disorders, and motor disorders. Paraesthesias are initially nocturnal and subsequently diurnal(2, 3). In advanced phases, weakness and thenar atrophy occur. The prevalence of CTS in the general population is about 3.4% in women and 0.6% in men(4). In addition to the case history and clinical examination, electrodiagnostic testing (EDT) is currently considered the gold standard for confirmation of a clinical diagnosis of CTS(5). Nevertheless, EDT is an expensive, time-consuming, and invasive test that is not readily accessible. In recent years, with the development and innovation of modern ultrasound (US) technology and high-frequency transducer, as well as continuously improved quality of US images, US technology is widely used in the diagnosis of CTS due to free invasion, simple operation, and no radiation(6).

US examination has unique advantages in the diagnosis of CTS, providing multiple reference indicators to improve diagnostic accuracy and reduce treatment risks(6). Lin et al.(7) demonstrates that US, particularly measurement of the median nerve cross-sectional area (CSA) at the carpal tunnel inlet, is a useful imaging modality for diagnosing CTS. Additional studies have shown that when the cross-sectional area is between 8.5 and 15.0 mm$^2$, there is a significant difference in sensitivity (62.0% to 97.9%) and specificity (63% to 100%) for the diagnosis of CTS(6). Besides median nerve swelling, sonoelastography and Doppler US reveal increased stiffness and vascularity of the nerve in CTS patients, providing additional information on disease activity and severity. Some scholars believe that the swelling ratio (SR) of the MN can be an essential indicator for the US diagnosis of CTS. The results of Sugimoto et al. showed that the critical value of the SR of the MN for the diagnosis of CTS is 1.55(8, 9). Wilson et al. believed that an SR of ≥1.3 had the highest sensitivity (72.5%) for the diagnosis of CTS. Buchberger et al. proposed the concept of the MN flattening ratio (FR), and their research results showed that the MN FR at the level of the hook of the hamate was significantly increased, with a critical diagnostic value of 4.2(10). Although many studies have reported the application of US in the diagnosis of CTS, there has always been controversy over the selection of different diagnostic parameters, which may be influenced by different diagnostic criteria, operator errors, measurement methods, machine resolution, and subjective differences among observers in delineating the MN. The following disputes have always existed: which index has the highest specificity and sensitivity? In which section is the diagnostic standard established?

Recent studies have shown that US has similar or even higher sensitivity and specificity than EDT in diagnosing CTS, as well as excellent intra- and inter-observer reliability(11). The reliability between evaluators is more likely to change, and a highly reliable US image processing method needs to be sought. Furthermore, currently, the tracking of MN in consecutive US images still depends on manual recognition and delineation, which requires massive human labor and makes it hard to implement the analysis clinically. Moreover, images acquired in dynamic US as the nerve is moving within the screen and are usually noisier than static images because of the motion artifacts, which further increases the difficulty in manual tracking.

To address these, an appealing solution is employing machine learning to segment the tracked nerve in the images automatically. In previous studies, deep learning has demonstrated satisfying performance in various applications(12-15). It has also been utilized to analyze the MN, while existing studies mostly focused on automatic tracking and segmentation of MN(16). Current methods predominantly adopt convolutional neural network (CNN) models(17), especially U-Net(18, 19), which counteracts the progressive loss of feature resolution with network depth through a symmetric U-shaped architecture combining low-level spatial information and high-level semantic features. Horng et al.(13) demonstrated the effectiveness of DeepNerve, a CNN-based model, for MN segmentation. They combined a modified U-Net, convolutional long short-term memory network, and MaskTrack method. However, their method requires manual labeling of the region of interest (ROI), limiting its flexibility. Festen et al.(19) utilized U-Net for MN segmentation and measurement. Similarly, Yang et al.(20) employed a modified Deeplabv3+ to segment carpal tunnel and its contents. Huang et al. and Shao et al.(21, 22)extended this work by incorporating attention mechanisms and modifying the encoder, respectively. However, Cosmo et al., Di Cosmo et al., and Smerilli et al. adopted a detection-based model, specifically Mask R-CNN, for MN segmentation(23-25). Recently, Yeh et al.(26) introduced Solov2-MN, a modified instance segmentation model, and reported modest improvements in segmentation accuracy. Notably, accurate segmentation in US images remains challenging. CNN-based methods often struggle to model relationships between distant elements due to the local inductive bias of convolutional operations. Although attention mechanisms and image pyramids have been attempted, they have not yielded significant improvements. In contrast, Transformers have emerged as an alternative architecture in computer vision, leveraging multi-head self-attention to effectively model long-range dependencies. The pioneering method, Vision Transformer (ViT)(27), exhibits high performance but computational complexity.

Despite their success, these methods could only provide the segmentation or bounding boxes of MN and failed to diagnosis the CTS directly. Recently, some attempts have been made to develop diagnostic tools for CTS. For example, Obuchowicz et al. utilized a feature-selection tool MaZda to find four key texture features to input into a Support Vector Machine (SVM) model(28). When conducted on 30 swollen MN and 30 normal people, the method achieved 79% accuracy on CTS diagnosis. Faeghi et al. developed a CAD system using radiomics features and SVM to diagnose CTS. The CAD system demonstrated higher performance than two radiologists, with an AUC of 0.926(29). For a more convenient diagnosis, Shinohara et al. used a deep learning algorithm to diagnose CTS directly from US images of the carpal tunnel inlet. They reported a high accuracy score, while this approach lacked clinical interpretability. However, an integrated diagnostic system that automatically calculates clinically quantifiable CTS diagnostic indicators to obtain more easily understandable classification results has not yet been developed.

In response to these clinical requirements, we propose a one-stop automated CTS

diagnosis system (OSA-CTSD) as an effective computer-aided diagnosis (CAD) tool to assist radiologists in diagnosing CTS using US images. To the best of our knowledge, the OSA-CTSD is the first method that combines three processes into a unified framework, including real-time MN delineation, accurate biometric measurements, and explainable CTS diagnosis. To comprehensively assess the clinal value of the method, the OSA-CTSD was evaluated on a large-scale US imaging dataset containing a total of 32,301 static images from US videos. It demonstrated promising diagnostic performance based on clinical-interpretable parameters, achieving the accuracy score of 93.85% and F1 score of 89.47%. Also, it was fully automated with a simplified scanning protocol (i.e., a straight sweep on the wrist). The utilization of such a powerful tool can not only reduce reliance on the expertise of examiners, but also has the potential to promote the future standardization of the CTS diagnosis process, benefiting both patients and radiologists.

**Methods**

*Study Participants*

This study was approved by the ethics committee of The Second People's Hospital of Shenzhen, The First Affiliated Hospital of Shenzhen University. Informed consent was obtained from all participants involved. Between May 2022 and January 2023, a total of 138 wrists were invited to participate in the study: 43 wrists from 31 patients with clinically evident carpal tunnel syndrome and typical clinical history, symptoms, and EDT(30); and 95 wrists from 52 healthy volunteers who met the inclusion criteria for age (at least 18 years). Exclusion criteria included contraindications for participation, previous wrist injury or surgery, central nervous system disorders, endocrine, metabolic, neuromuscular, musculoskeletal disorders relevant to CTS development, bifid MN, or persistent median artery. Three wrists from two patients and five wrists from three healthy volunteers were excluded based on these criteria.

The final sample consisted of two groups: a CTS group comprising 40 wrists in 29 patients (mean age =55.1 years; range=33-72 years), and a control group comprising

90 wrists in 52 healthy volunteers (mean age=29.4 years; range=22-67 years). (Fig 1)

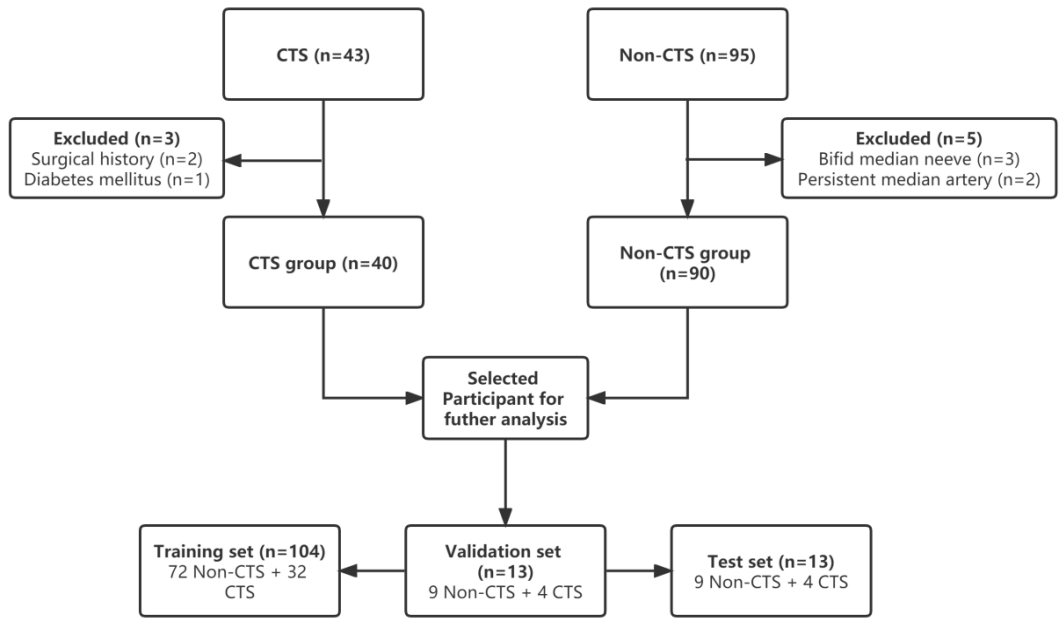

Figure 1. Selection of CTS (wrists from CTS patients) and Non-CTS (wrists from healthy volunteers). According to access criteria, 40 wrists from 29 CTS patients and 90 wrists in 52 healthy volunteers were included. In total, 130 videos were divided into 104, 13, and 13 for the training, validation, and internal test sets, respectively.

*US Technique*

The study utilized a PHILIPS EPIQ 7C US platform with a non-linear transducer that had an acquisition frequency of 5-12 MHz to image the subjects. The subjects were seated facing the examiner, with their arms extended and wrists rested on a hard flat surface while their forearms were supinated. To obtain dysnamic video images of the MN axial B-mode, a straight sweep is performed from about 2.5cm proximal to the flexor support band as the starting point to 2.5cm distal as the endpoint, including the proximal end of the wrist canal, entrance, and exit. During the entire scanning process, only minimal deviation (<1cm) between the probe center and palmar midpoint is required by US radiologists (to ensure that MN is collected), without imposing special restrictions on imaging parameters (e.g., depth, etc.) for better reproducibility in the future applications. This results in depths ranging from 3.0cm to 4.0cm, frame rates ranging from 45Hz to 51Hz. Each video has an average duration of 4.97 seconds, consisting of 248 frames.

The study included patients exhibiting clinical symptoms of CTS, such as numbness and pain in fingers supplied by MN along with the decline in some active and passive motor functions as study subjects, while healthy individuals without any clinical symptoms served as the control group. A total of 130 dynamic videos were randomized for independent viewing and diagnosis by three experienced radiologists having more than ten years of experience in musculoskeletal US diagnostics along with three inexperienced radiologists who underwent a two-hour training session but had no prior exposure to musculoskeletal US diagnostics.

All radiologists remained blinded to both clinical diagnosis results as well as diagnostic results provided by the other five radiologists throughout this process eliminating potential subjective interference factors. Finally, two experienced musculoskeletal US diagnosticians used internal software to describe all dynamic video images obtained during this process. (Fig. 2)

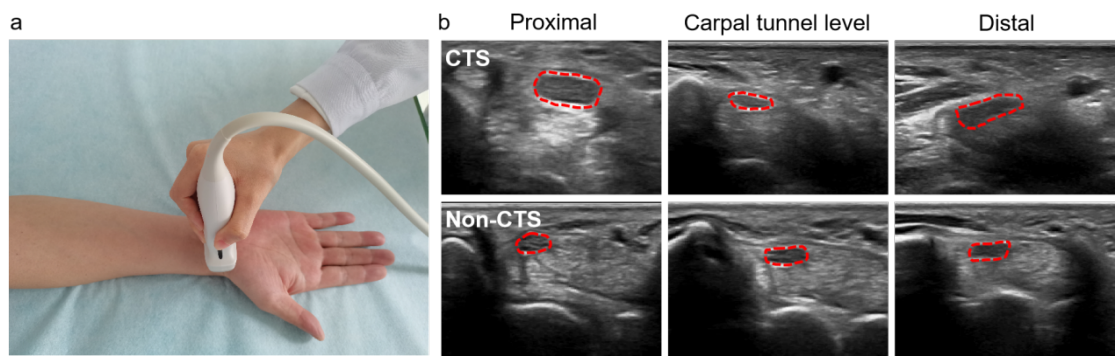

Figure 2. (a) The probe was positioned just proximal to the carpal tunnel inlet, with their arms extended and wrists rested on a hard flat surface. (b) US images of CTS patients and healthy volunteers at the proximal, entrance, and exit locations of the carpal tunnel.

*Dataset*

In total 81 participants were enrolled in this study (52 normal participants and 29

patients). 130 videos were obtained following the image collection protocol. A total of 32,301 2D images were extracted from the raw videos, without any down-sampling or interpolation. To the best of our knowledge, it is currently the largest MN US dataset. The whole dataset is randomly split at the patient level into 8:1:1, for training (25,326 images), validation (3,527 images), and testing (3,448 images), respectively. The only image pre-processing we carried out is anonymization to remove privacy information and ensure reliable performance in handling unseen test cases in real-world applications. All these US images were delineated by experienced radiologists using Pair software (http://www.aipair.com.cn/). The ground truth (GT) MN measurements of each patient were then calculated based on these delineations. The GT diagnostic labels were confirmed by an experienced radiologist (>8-year experience) based on the combination of EDT results and clinical symptoms.

*Model design*

To facilitate a time-efficient, labor-saving diagnosis process, we propose a one-stop automated CTS diagnosis system (OSA-CTSD) that integrates real-time MN delineation, accurate biometric measurements, and explainable CTS diagnosis into a unified framework. It can process raw US images and does not require any human intervention. Figure 3 displays the overall schematic.

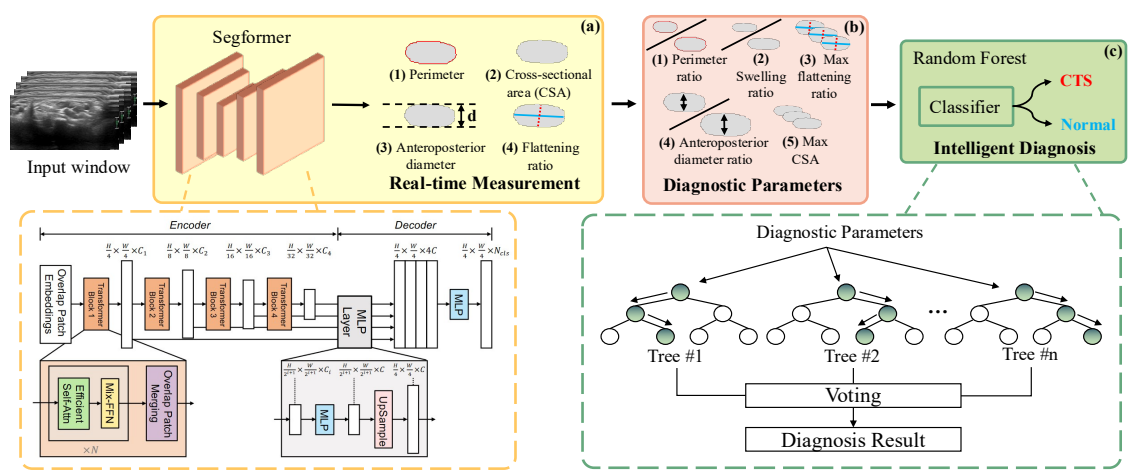

Figure 3. The proposed method (OSA-CTSD) used in the present study mainly consists of three processes: real-time MN delineation, accurate biometric measurements, and explainable CTS diagnosis.

**Segmentation model** Segmentation is a vital technique that involves locating and delineating MN for analysis or visualization. Accurate and consistent MN delineation is the fundamental step in the diagnosis of CTS, as it provides morphological characteristics of MN and helps the interpretation of the severity of the condition. Compared to the CNN-based models and ViT, SegFormer(31) introduced a more simple and more efficient Transformer-based architecture. As shown in Figure 3, it consists of two primary components: a positional-encoding-free, hierarchically structured Transformer encoder and a lightweight All-MLP decoder. An input image of size H×W×3 is divided into patches of size 4×4, which differs from the coarse-grained patch approach used in ViT. The patches are then fed to the hierarchically structured Transformer encoder, which produces multi-level features with different resolutions. The resulting features are directed to the All-MLP decoder, which predicts the segmentation mask. The SegFormer model has six variants ranging from SegFormer-B0 to SegFormer-B5, each with varying hyperparameters of the encoder (e.g., $K_i$: patch size, $S_i$: stride, $P_i$: padding size, $C_i$: channel number, $L_i$: number of encoder layers of Stage $i$, etc.), while maintaining the same model design. For example, the B0 is the most compact one, while B5 has the largest modeling capacity. In this study, we opted for the SegFormer-B2 variant to achieve a balanced trade-off between model performance and computational cost. For more information on the specific hyperparameters used, please refer to the original paper(31).

**Automated measurement** In the automated measurement of MN masks, segmentation helps to precisely identify and measure the nerve structure from complex medical images. To translate the segmentation results into diagnosis-related descriptors, biometric measurements are commonly used to monitor the changes in size, shape, and location of the MN. It is helpful to clarify the diagnosis and to provide guidance for the choice of subsequent treatment. Existing approaches either require manual selection of ROI or are restricted by unstable scanning location or inconsecutive scanning window, while the proposed OSA-CTSD system is intervention-free and utilizes a unified scanning protocol. In specific, we first measure the morphological parameters of the MN in each 2D image of the same patient. Fig. 3a depicts the automated measurement process, where morphological parameters acquired from MN in each image are perimeter (Fig. 3a (1)), cross-sectional area (CSA) (Fig. 3a (2)), anteroposterior diameter (AD) (Fig. 3a (3)), flattening ratio (FR) (Fig. 3a (4)), where FR is defined as the ratio of horizontal to vertical diameter of the nerve.

As the level of swelling and area variation are strong indicators of CTS, we calculate a set of diagnostic measurements to better summarizes the overall MN status of a patient based on all obtained 2D parameters. This set consists of the perimeter ratio (PR) (Fig. 3b (1)), swelling ratio (SR) (Fig. 3b (2)), maximum flattening ratio (Max FR) (Fig. 3b (3)), anteroposterior diameter ratio (ADR) (Fig. 3b (4)), and maximum cross-sectional area (Max CSA) (Fig. 3b (5)), which form integral components of our approach. The definition of PR and ADR (Fig. 3b (1), b (3)) is the ratio of maximum and minimum values of respective parameters acquired from Fig. 3a. And SR (Fig. 3b (2)) is defined as the ratio of CSA at MN compression to CSA at swelling.

**Explainable diagnosis** EDT and physical examination are common diagnostic tests for CTS, while they could introduce patient discomfort, inconclusive results, and false-positive diagnoses. US is immune from these limitations, and its diagnostic results correlate well with EDT and clinical examination results. However, performing accurate diagnoses based on US relies on experience and skills. The CAD system can assist this effectively. To fully exploit the MN descriptors defined in the previous section, we propose to use random forest due to its additional insights into feature importance and interactions, facilitating model interpretation and feature selection. Our random forest model consisted of 135 trees, with a maximum depth per tree set to 6. To prevent overfitting, we included a minimum of 12 samples required for splitting internal nodes and a minimum of 9 samples to form a new leaf node. We also selected a maximum of log2 features to consider at each split to increase the diversity among trees.

The model's performance was measured using the Gini index, and the random state was set to 90 for reproducibility.

*Experimental setup*

**Experiments** To fully evaluate the proposed method, we conducted comparative experiments in three key stages, namely segmentation, measurement, and diagnosis. All the experiments were evaluated on our in-house MN dataset. 1). We implemented several state-of-the-art (SOTA) methods in segmenting the MN. (i.e., U-Net-MN(19), DeepLabV3+-MN(20), Mask R-CNN-MN(25), and Solov2-MN(26)). In order to explore the potential of these models, we selected their best performance variants in the comparison (i.e., DeepLabV3+-MN, Mask R-CNN-MN, and Solov2-MN with the backbone of ResNet-101)(26, 32). Also, for simplicity and fairness, these models were trained suitably according to their best hyper-parameter selection. 2). To ensure the automated segmentation results could generate biometric measurements that were consistent with the manual results, we calculated both frame-level measurements (i.e., perimeter, CSA, ADR, and FR), and video-level measurements (PR, SR, Max CSA, ADR, and Max CSA) from the automated delineations. This extension not only further examined the segmentation capability of the models but also provided a solid foundation for subsequent diagnosis. 3). Finally, we investigated whether OSA-CTSD could benefit the diagnosis process by conducting the following reader study. In specific, we invited two groups of radiologists with different levels of experience to classify the MN US videos. Each of the groups contained 3 radiologists to provide statistically consistent results. To further examine the robustness of the model, we also implemented OSA-CTSD variants with Logistic regression and SVM classifiers, respectively. Note that the rest of the framework maintained the same for a fair comparison.

**Evaluation Metrics** The segmentation performance was evaluated based on Intersection over Union (IoU), Dice coefficient (Dice), HD95 (95th percentile of Hausdorff Distance), and ASSD (Average Symmetric Surface Distance). IoU and Dice measure the overlap between the prediction and the GT.

HD95 measures the distance between two sets of points (typically, the points on the boundaries of two segmented regions). Suppose that $A$ and $B$ represent two sets of points, where $A$ is the GT and $B$ is the segmentation result. Then the HD95 score is calculated as follows:

$$HD95(A, B) = max\big(h_{95}(A, B), h_{95}(B, A)\big),$$

where $h_{95}(A, B)$ is the 95th percentile of the Hausdorff distance between points in set $A$ and points in set $B$, and $h_{95}(B, A)$ is the 95th percentile of the Hausdorff distance between points in set $B$ and points in set $A$.

The ASSD score measures the average distance between the surfaces of two segmented regions, and a smaller value of ASSD indicates a better segmentation performance. The ASSD score is calculated as follows:

$$ASSD(A, B) = \frac{1}{|A|+|B|}\big(\sum_{a \in A} d(a, B) + \sum_{b \in B} d(b, A)\big),$$

where $|A|$ and $|B|$ are the numbers of points in set $A$ and set $B$, respectively, and $d(a, B)$ and $d(b, A)$ are the shortest distances from point $a$ in set $A$ to set $B$ and from point $b$ in set $B$ to set $A$, respectively.

Based on segmentation results, we evaluated the measurement performances of different segmentation models by mean absolute error (MAE) in both levels. Qualitative analyses were also conducted to evaluate segmentation and measurement results visually. Meanwhile, the classification performance was evaluated using accuracy (ACC), sensitivity (SEN), specificity (SPE), F1 score (F1), receiver operating characteristic curve (ROC), area under the curve (AUC), false negative rate (FNR), and false positive rate (FPR).

**Statistical Analysis** We also conducted statistical analysis to fully demonstrate the superiority of the proposed method. In specific, the differences in segmentation metrics (Dice, IoU, HD95, ASSD) among models were compared using Wilcoxon signed-rank test. Furthermore, measurement (e.g., perimeter, CSA, PR, SR, etc.) evaluation metric (Absolute Error) was also analyzed by Wilcoxon signed-rank test on two levels (frame-level and video-level). For the reader study, chi-square test was used to identify the statistical differences in diagnostic performance among two groups of radiologists and that of the proposed system. Also, the statistical significances of the variants of proposed system were examined by Delong's test. For each statistical test, $p<0.05$ was considered to indicate significance. The analysis was performed using software package SciPy statistical toolkit in Python version = 3.8.

**Implementation details** Our proposed method was trained and evaluated on an Nvidia GeForce RTX 3090 GPU. The method was performed using the Pytorch, OpenCV, and Scikit-learn packages. Our segmentation neural network used is SegFormer-B2, a variant with encoder parameters number of 24.2M that balances segmentation performance and inference speed. We conducted data augmentation via random horizontal flipping, random resize with a ratio of 0.5-2.0, and random cropping to dimensions of 512×512. Our training procedure involves using the AdamW optimizer for 160K iterations on the MN dataset, and the batch size was 16. We set the learning rate to an initial value of 1e-6 and then apply a "poly" learning rate schedule with factor 1.0 by default. We implemented all models and conducted hyper-parameters selection for all experiments suitably. Meanwhile, the classifiers were trained on diagnostic parameters calculated from radiologists' delineations and tested on diagnostic parameters calculated from our segmentation.

**Results**

*MN segmentation results*

We compared some popular CNN-based semantic segmentation models (U-Net-MN, DeepLabV3+-MN), instance segmentation models (Mask R-CNN-MN, Solov2-MN), and our model to delineate the MN region. Quantitative segmentation evaluation results were displayed in Table 1. Among these models, ours exhibited the highest Dice and IoU values, indicating that its predictions were more similar to the GT ($p<0.05$). Additionally, ours demonstrated the lowest HD95 and ASSD values, indicating that its boundaries correlated better with the manual results ($p<0.05$). In contrast, U-Net-MN showed the lowest Dice and IoU values and the highest HD95 and ASSD values, suggesting that its segmentation performance was relatively poor (with statistically significant difference, $p<0.05$). Additionally, because U-Net-MN used the original U-Shape architecture, its capability of feature extraction was not comparable to the others and performed worst. The heavy encoders used by DeepLabV3+-MN, Mask R-CNN-MN, and Solov2-MN led to good results, even DeepLabV3+-MN achieved comparable results to ours (with no statistically significant differences in ASSD, Dice, IoU; $p$-values of 0.13, 0.54, 0.54 respectively). However, they also brought a larger number of parameters and slower inference speeds. It is worth mentioning that the proposed OSA-CTSD used only 1/3 the number of parameters of that used by DeepLabV3+-MN, Mask R-CNN-MN, and Solov2-MN, while achieving superior performance (see Table 1).

Table 1. Segmentation performances of U-Net-MN, DeepLabV3+-MN, Mask R-CNN-MN, Solov2-MN, and ours. Parameters number of whole model and test speed are also present in the table. ("↓" indicates that the smaller the value, the better the performance. "↑" indicates the larger the value, the better the performance.)

| Method | HD95 (px) ↓ | $p$ | ASSD (px) ↓ | $p$ | Params (M) ↓ |
|---|---|---|---|---|---|
| U-Net-MN | 8.44±6.38 | <0.05 | 3.08±2.38 | <0.05 | 29.06 |
| DeepLabV3+-MN | 7.51±4.87 | <0.05 | 2.69±1.79 | 0.13 | 62.57 |
| Mask R-CNN-MN | 8.52±6.47 | <0.05 | 2.90±2.03 | <0.05 | 62.74 |
| Solov2-MN | 7.64±5.33 | <0.05 | 2.79±2.02 | <0.05 | 65.22 |
| Ours | 7.21±4.74 | - | 2.64±1.52 | - | 24.73 |
| Method | Dice (%) ↑ | $p$ | IoU (%) ↑ | $p$ | FPS (img/s) ↑ |
| U-Net-MN | 84.54±10.85 | <0.05 | 74.50±13.84 | <0.05 | 35.81 |
| DeepLabV3+-MN | 85.76±8.59 | 0.54 | 75.93±11.75 | 0.54 | 28.72 |
| Mask R-CNN-MN | 84.80±9.69 | <0.05 | 74.74±12.69 | <0.05 | 27.44 |
| Solov2-MN | 85.20±9.85 | <0.05 | 75.27±12.53 | <0.05 | 28.75 |
| Ours | 85.78±8.71 | - | 76.00±11.98 | - | 47.28 |

*HD95* Hausdorff distance 95th percentile; *ASSD* Average symmetric surface distance; *Dice* Dice

coefficient; *IoU* Intersection over union; *Params* Parameters; *FPS* Frames per second; *px* pixels; *img/s* Images per second; *M* Million.

The qualitative segmentation results of our method and competing methods were also present in the Figure 4. In general, ours as well as DeepLabV3+-MN performed well, and U-Net-MN, Mask R-CNN-MN, and Solov2-MN obtained bad performances occasionally. For example, although there was no significant variance among these models in frame 185, the segmentation performance degradation could be obviously observed in frame 219, where U-Net-MN failed to recognize MN at all. Also note that only ours and DeepLabV3+-MN delineated the left and right boundaries of MN in frame 219 and achieved high segmentation accuracy. Meanwhile, the segmentation results of Mask R-CNN-MN and Solov2-MN both deviated from the GTs in the upper regions. And the left and right MN boundaries obtained from instance segmentation models (i.e., Mask R-CNN-MN, Solov2-MN) displayed discrepancies compared to the GTs in frames 219 and 254. Also, the segmentation area of Solov2-MN was smaller than the GT in frame 254. Frame 307 represented the end of the US scanning, which also meant that the MN had reached the deepest point in the arm. Thus, all the models were not performing well: the delineation of U-Net-MN outlined in the lower left of the MN was too thick, but it performed better at the right nerve boundary. DeepLabV3+-MN performed poorly at the right boundary, which was far from the accurate morphology of MN. While the performance of Mask R-CNN-MN was not as poor as the two models, its segmented morphology still considerably differed from the GT. Solov2-MN yielded the poorest performance in frame 307, as it erroneously identified the MN as a stripe-like object.

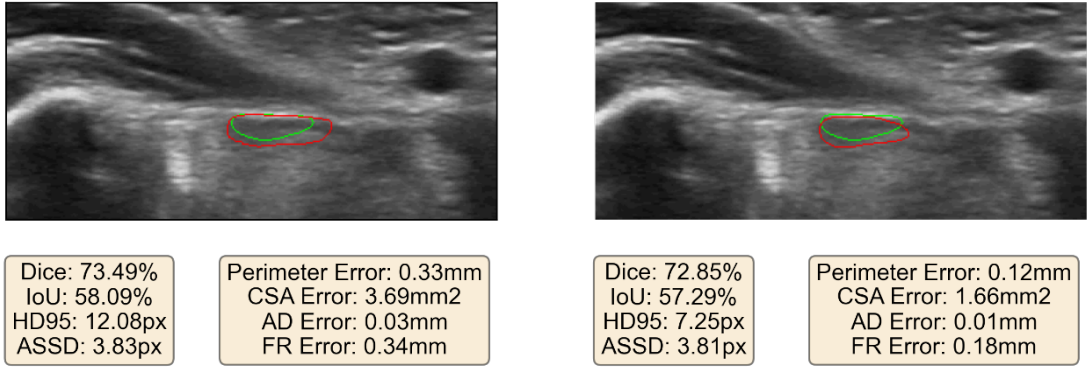

Figure 4. Visual samples of the segmentation result delineated by different segmentation models. "GT" and "Ours" refer to the GT and SegFormer-B2.

*MN automated measurement results*

To evaluate segmentation performance more comprehensively, we extended the experiment to frame-level and video-level measurements of MN. MAEs of MN morphological measurements (frame-level) and diagnostic parameters (video-level) were shown in Table 2 and Table 3, respectively. As we can observe in Table 2, the proposed produced the smallest MAE values among all competing methods. Our approach outperformed its counterparts in all of the parameters measurement tasks (with statistically significant differences, $p<0.05$). As the frame-level morphological parameters were directly acquired from the corresponding segmentation predictions, their measurement performance of each model aligned well with that shown in Table 1. For example, U-Net-MN achieved relatively larger measurement errors (e.g., $1.67\pm0.60$ vs $1.41\pm0.44$ and $2.30\pm1.21$ vs $1.71\pm0.80$, $p<0.05$) with the lowest Dice and IoU scores, and DeepLabV3+-MN performed relatively better in measuring CSA and AD. However, there were a few exceptions as could be observed in Table 2. For example, the MAEs of the perimeter, CSA, and AD acquired by DeepLabV3+-MN were slightly higher than those of Solov2-MN, while its FR MAE score was lower than the latter. Additionally, Mask R-CNN-MN exhibited another exception where their perimeter and FR MAE scores were higher despite having a better Dice score. An explanation is that measurement results are not decided directly by segmentation metrics, but these metrics are correlated. Also note that the MAE values of different methods were relatively close, as they exhibited the averaged deviation of perimeter, CSA, AD, and FR across all frames of all test videos.

Table 2. MN morphological parameters measurement performances of U-Net-MN, DeepLabV3+-MN, Mask R-CNN-MN, Solov2-MN, and ours evaluated by MAE. ("↓" indicates the smaller the value, the better the measurement performance.)

| Method | Mean Absolute Error (MAE) ↓ | | | |
| --- | --- | --- | --- | --- |
| | Perimeter (mm) | CSA (mm2) | AD (mm) | FR (%) |
| U-Net-MN | 1.67±0.60 | 2.30±1.21 | 0.28±0.09 | 35.51±14.47 |
| DeepLabV3+-MN | 1.44±0.45 | 1.80±0.80 | 0.21±0.09 | 30.94±7.73 |
| Mask R-CNN-MN | 1.69±0.85 | 1.77±0.93 | 0.22±0.11 | 39.32±14.80 |
| Solov2-MN | 1.43±0.38 | 1.69±0.76 | 0.20±0.08 | 34.40±14.10 |
| Ours | 1.41±0.44 | 1.71±0.80 | 0.20±0.09 | 30.65±9.61 |

*CSA* Cross-sectional area; *AD* Anteroposterior diameter; *FR* Flattening ratio.

On the contrary, the video-level measurements are often defined by the maximum difference of the MN across frames in a single video. Therefore, a few outliers in frame-level measurements could lead to large deviations in video-level ones. Table 3 displayed the video-level results. For example, U-Net-MN showed 262.95% deviation in PR and 207.44% deviation in ADR (see row 1, Table 3). After examining its predictions, we found that this method failed to identify the MN or delineate the correct shape of the MN in some frames, which led to considerably higher PR and ADR MAE values ($p<0.05$). They adopted a symmetric structure that may cause information bottleneck problems, which cannot make full use of contextual information and local detail information. Note that the differences among competing methods reported in Table 3 are much higher than that in Table 2. As stated earlier, the video-level measurement values may deteriorate due to the increase of outliers, while the frame-level measurements may smooth out these exceptions.

Table 3. MN automated diagnostic parameters measurement performances of U-Net-MN, DeepLabV3+-MN, Mask R-CNN-MN, Solov2-MN, and ours evaluated by MAE. ("↓" indicates the smaller the value, the better the measurement performance.)

| Method | Mean Absolute Error (MAE) ↓ | | | | |
|---|---|---|---|---|---|
| | PR (%) | SR (%) | ADR (%) | Max FR (%) | Max CSA (mm2) |
| U-Net-MN | 262.95 ±434.85 | 59.38 ±63.04 | 207.44 ±367.47 | 44.89 ±35.44 | 2.61±1.70 |
| DeepLabV3+-MN | 27.13 ±33.36 | 58.72 ±118.58 | 14.26 ±11.66 | 33.40 ±29.34 | 1.84±1.33 |
| Mask R-CNN-MN | 23.32 ±28.34 | 54.56 ±64.02 | 15.78 ±15.98 | 37.13 ±27.98 | 1.80±1.35 |
| Solov2-MN | 31.06 ±49.94 | 47.95 ±57.53 | 14.78 ±14.16 | 51.95 ±38.22 | 2.89±3.29 |
| Ours | 19.57 ±28.44 | 56.01 ±112.15 | 14.48 ±15.22 | 36.08 ±36.43 | 1.82±2.27 |

*PR* Perimeter ratio; *SR* Swelling ratio; *ADR* Anteroposterior diameter ratio; *Max FR* Maximum flattening ratio; *Max CSA* Maximum cross-sectional area.

*Carpal tunnel syndrome diagnosis results*

Table 4 reported the quantitative results of our reader study that evaluated the diagnostic accuracy for CTS among radiologists with varying levels of experience, including three inexperienced radiologists (L, P, Z) and three experienced radiologists (C, G, W). The experienced radiologists outperformed the inexperienced radiologists in terms of all metrics (with the higher ACC, SEN, SPE, F1, and the lower FNR, FPR) with large margins and statistically significant differences ($p<0.05$). They scored an average ACC of 98.21% - around 8% higher than that of the inexperienced group. Furthermore, although the average SPE is only about 4% higher than that of less experienced radiologists, the SEN is as much as 16% higher. This not only results in a high F1 score (97.08%) but also in very low FNR and FPR (3.33% and 1.11%, respectively). Meanwhile, among the group of inexperienced radiologists, there was a difference of up to 10% in SEN between different individuals. It was also noted that the average SEN score of inexperienced radiologists was about 10% lower than their SPE score (80.83% vs 94.44%).

Table 4. Diagnosis performances of radiologists with varying levels of experience and different variants of our proposed method (OSA-CTSD). "Ours+LR" and "Ours+SVM" refer to OSA-CTSD variants using and Logistic regression and SVM classifier, respectively. ("↓" indicates that the smaller the value, the better the performance. "↑" indicates the larger the value, the better the performance.)

| Radiologists | | ACC (%) ↑ | SEN (%) ↑ | SPE (%) ↑ | F1(%) ↑ | FNR (%) ↓ | FPR (%) ↓ |
|---|---|---|---|---|---|---|---|
| Inexperienced | L | 93.08 | 85.00 | 96.67 | 88.31 | 15.00 | 3.33 |
| | P | 90.77 | 82.50 | 94.44 | 84.62 | 17.50 | 5.56 |
| | Z | 86.92 | 75.00 | 92.22 | 77.92 | 25.00 | 7.78 |
| Avg | | 90.26 ±3.11 | 80.83 ±5.20 | 94.44 ±2.22 | 83.62 ±5.27 | 19.17 ±5.20 | 5.56 ±2.23 |
| Experienced | C | 98.46 | 97.50 | 98.89 | 97.50 | 2.50 | 1.11 |
| | G | 98.46 | 95.00 | 100.00 | 97.44 | 5.00 | 0.00 |
| | W | 97.69 | 97.50 | 97.78 | 96.30 | 2.50 | 2.22 |
| Avg | | 98.21 ±0.44 | 96.67 ±1.44 | 98.89 ±1.11 | 97.08 ±0.68 | 3.33 ±1.44 | 1.11 ±1.11 |

| Model variants | ACC (%) ↑ | SEN (%) ↑ | SPE (%) ↑ | F1 ↑ (%) | FNR (%) ↑ | FPR (%) ↑ |
|---|---|---|---|---|---|---|
| Ours+LR | 92.31 | 80.00 | 97.78 | 86.49 | 20.00 | 2.22 |
| Ours+SVM | 93.08 | 82.50 | 97.78 | 88.00 | 17.50 | 2.22 |
| Ours | 93.85 | 85.00 | 97.78 | 89.47 | 15.00 | 2.22 |

*ACC* Accuracy; *SEN* Sensitivity; *SPE* Specificity; *F1* F1 score; *FNR* False negative rate; *FPR* False positive rate.

In addition, we implemented several variants of our proposed models by replacing different classifiers to evaluate the robustness of our models. Results showed that these variants scored slightly lower ACC, and F1 scores than OSA-CTSD, while still outperforming than those of the inexperienced radiologists ($p<0.05$). There were no significant differences among these variants ($p>0.05$), which demonstrated that our proposed framework was general and compatible with various classifiers. Moreover, to further investigate the influence of individual diagnostic parameters, we implemented 5 LR models using only PR, SR, ADR, Max FR, and Max CSA, respectively. Their diagnostic performances were reported in Table 5. The corresponding ROC curves were plotted in Fig. 5, which is a plot of the true-positive rate and false-positive rate on the coordinate axis. Our approach presented a robust classification of CTS, with sensitivity of 85.00%, specificity of 97.78%, accuracy of 93.85%, F1 score of 89.47%, FNR of 15.00%, FPR of 2.22%, and AUC of 0.98.

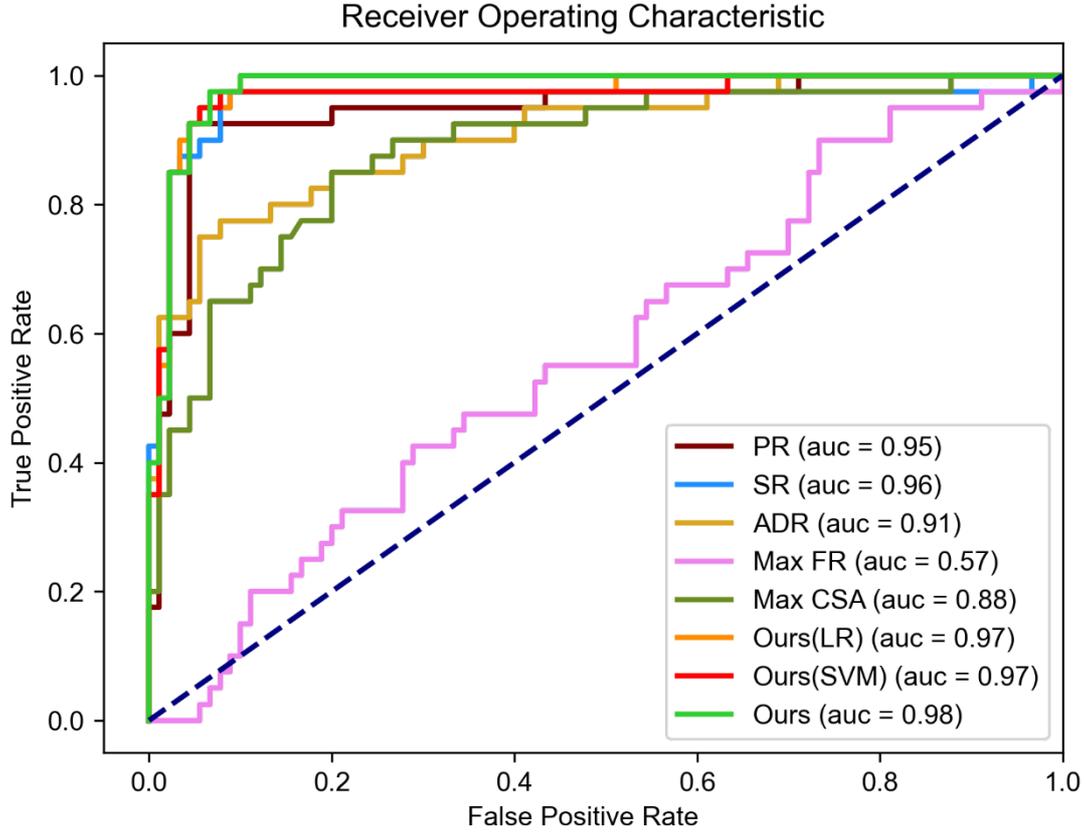

Figure 5. Receiver operating characteristic curve of OSA-CTSD and LR models using only PR, SR, ADR, Max FR, and Max CSA, respectively. *PR* Perimeter ratio; *SR* Swelling ratio; *ADR* Anteroposterior diameter ratio; *Max FR* Maximum flattening ratio; *Max CSA* Maximum cross-sectional area; *LR* Logistic regression; *SVM* Support vector machine.

**Discussion**

The quantitative segmentation results of the same MN dataset shows that ours achieved the most balanced and powerful performance among these models. While Dice and IoU are important metrics, the ASSD and HD95 are more sensitive to the following biometric measurements. As we can see in Table 3, Mask R-CNN-MN and Solov2-MN produce relatively poorer MAE values, while their Dice and IoU scores are still acceptable (Table 1). It may stem from that their HD95 and ASSD are considerably larger and may lead to inaccurate measurement results. In Figure 6, there were two segmentation results from different models. It could be observed that the higher HD95 scores could result in worse measurement results when their Dice and IoU scores are close. In our study, we found that HD95 and ASSD results are more closely related to measurement results (such as PR, CSA, AD, and FR) than other metrics. It is because these metrics characterize the boundary of the MN, whereas Dice and IoU scores focus more on the degree of overlap between the predictions and the GTs. We conjecture that the relatively poorer performance of Mask R-CNN-MN and Solov2-MN in measuring PR and SR may be explained by that they both utilized instance segmentation models. These approaches are known to be effective in detecting a large number or occlusion of objects in an image, while may not be well-suited for the task of contouring the MN in ultrasonography due to the lack of multi-scale skip connections and detailed boundary information. Also, the misaligned and low-resolution mask head of Mask R-CNN-MN could lead to inaccurate and blurry edge segmentation. Meanwhile, Solov2-MN also used the dynamic convolution kernel, which is highly sensitive to the quality and resolution of the input features. As a result, their approaches resulted in higher HD95 scores and performed poorly in video-level measurements when applied to our dataset.

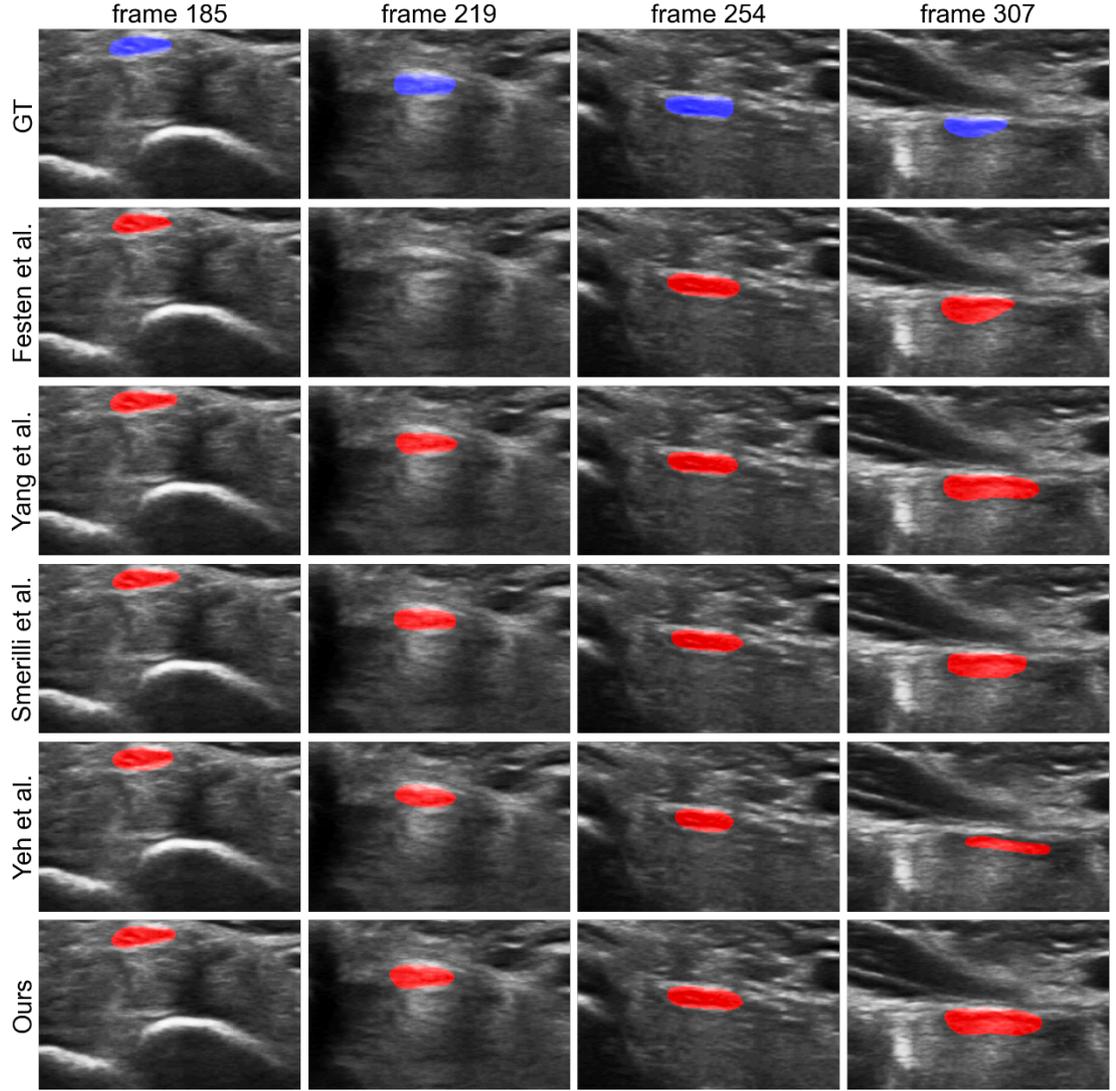

Figure 6. Visual samples of the correlation between the different segmentation scores and measurement errors. Red regions refer to the segmentation result, green regions refer to the GT. *Dice* Dice coefficient; *IoU* Intersection over union; *HD95* Hausdorff distance 95th percentile; *ASSD* Average symmetric surface distance; *CSA* Cross-sectional area; *AD* Anteroposterior diameter; *FR* Flattening ratio.

In addition, OSA-CTSD is not only reliable in segmentation but also is computation efficient. The inference speed (the number of inferred frames per second) of it can reach 47.2 FPS, which is the fastest of these models (Table 1). It may be explained by that our model utilized a lightweight All-MLP decoder that could capture powerful representations with a limited number of parameters. Also due to the adoption of the hierarchically structured Transformer encoder for ours, this allows the models to better capture context information, which is essential for accurate segmentation. It is worth pointing out that in the last column in Fig. 4 that all methods have difficulty in identifying the MN boundaries (especially the right). As scanning goes further and MN's position gets deeper, the boundary of the MN becomes unclear, bringing additional challenges in capturing the target. Note that unstable performance in such a scenario could lead to inaccurate frame-level measurements and ultimately hamper video-level measurement accuracy. Therefore, accurate hard-case segmentation such as those generated by OSA-CTSD builds a firm foundation for subsequent measurement and diagnosis.

As reported in Table 4, the experienced group have showed performance to inexperienced one in diagnostic evaluation metrics (e.g., ACC, SEN, SPE, etc.), especially higher F1 score. These results indicate that experienced radiologists are not only capable of accurately diagnosing CTS, but also are better at balancing the FNR and FPR, thus reducing missed and erroneous diagnoses. However, due to the shortage of experienced radiologists, timely and accurate diagnosis of CTS may not always be possible, leading to delayed or inappropriate treatment for patients. Meanwhile, the lower SEN score and higher SPE score of inexperienced radiologists suggest a potential conservative bias in their diagnostic tendencies, i.e., inclining to identify Non-CTS people. This could lead to a higher FNR and more missed diagnoses.

Table 5. Diagnosis performances of Logistic regression (LR) models using only PR, SR, ADR, Max FR, and Max CSA, respectively.

| Method | ACC (%) ↑ | SEN (%) ↑ | SPE (%) ↑ | F1 ↑ (%) | FNR (%) ↓ | FPR (%) ↓ |
|---|---|---|---|---|---|---|
| PR | 80.00 | 37.50 | 98.89 | 53.57 | 62.50 | 1.11 |
| SR | 92.31 | 80.00 | 97.78 | 86.49 | 20.00 | 2.22 |
| ADR | 80.77 | 40.00 | 98.89 | 56.14 | 60.00 | 1.11 |
| Max FR | 69.23 | 0.00 | 100.00 | 0.00 | 100.00 | 0.00 |
| Max CSA | 80.77 | 50.00 | 94.44 | 61.54 | 50.00 | 5.56 |
| Ours | 93.85 | 85.00 | 97.78 | 89.47 | 15.00 | 2.22 |

*PR* Perimeter ratio; *SR* Swelling ratio; *ADR* Anteroposterior diameter ratio; *Max FR* Maximum flattening ratio; *Max CSA* Maximum cross-sectional area.

To avoid missed diagnoses as well as enhance diagnostic efficiency, there is a need for a convenient, efficient, quantitative, and interpretable automated diagnostic tool. The proposed fully automated OSA-CTSD offers such a solution. Notably, OSA-CTSD could delineate the MN in real-time (Table 1), calculate diagnostic parameters for CTS, and rapidly generate accurate diagnosis results without any manual intervention. In specific, the SEN and F1 scores of OSA-CTSD were significantly higher than the average levels of the inexperienced group with statistically significant difference ($p<0.05$). It can be seen that it reported better performance than the inexperienced group in all metrics. In terms of single factor analysis (Table 5), the superiority of the OSA-CTSD may be explained by that it fully incorporates comprehensive MN structural information regarding the proximal entrance, carpal tunnel segment, and distal exit segments to analyze the MN. If only measuring maximum CSA alone, some patients may be missed due to less severe swelling near proximal or distal ends despite clear compression at the carpal tunnel site; however, using SR as an auxiliary diagnosis may yield better results. This suggests that the OSA-CTSD may help to reduce the number of missed diagnoses. On the other hand, OSA-CTSD scored similar SPE and FPR score with that of the experienced radiologists and even reached the level of one of the experienced radiologists (W). It was not regarded as statistically significant difference in the diagnostic efficacy between the experienced group and OSA-CTSD ($p=0.08$). These results suggest that OSA-CTSD demonstrates significantly superior diagnostic performance compared to inexperienced group and is also more closely aligned with that of experienced radiologists. For inexperienced radiologists, OSA-CTSD can not only provide a comprehensive measurement report of the MN, but also diagnostic suggestions to better assist their diagnostic workflow. Meanwhile, OSA-CTSD may also help the experienced radiologists by automating the measuring process of the MN that can accelerate the evaluation.

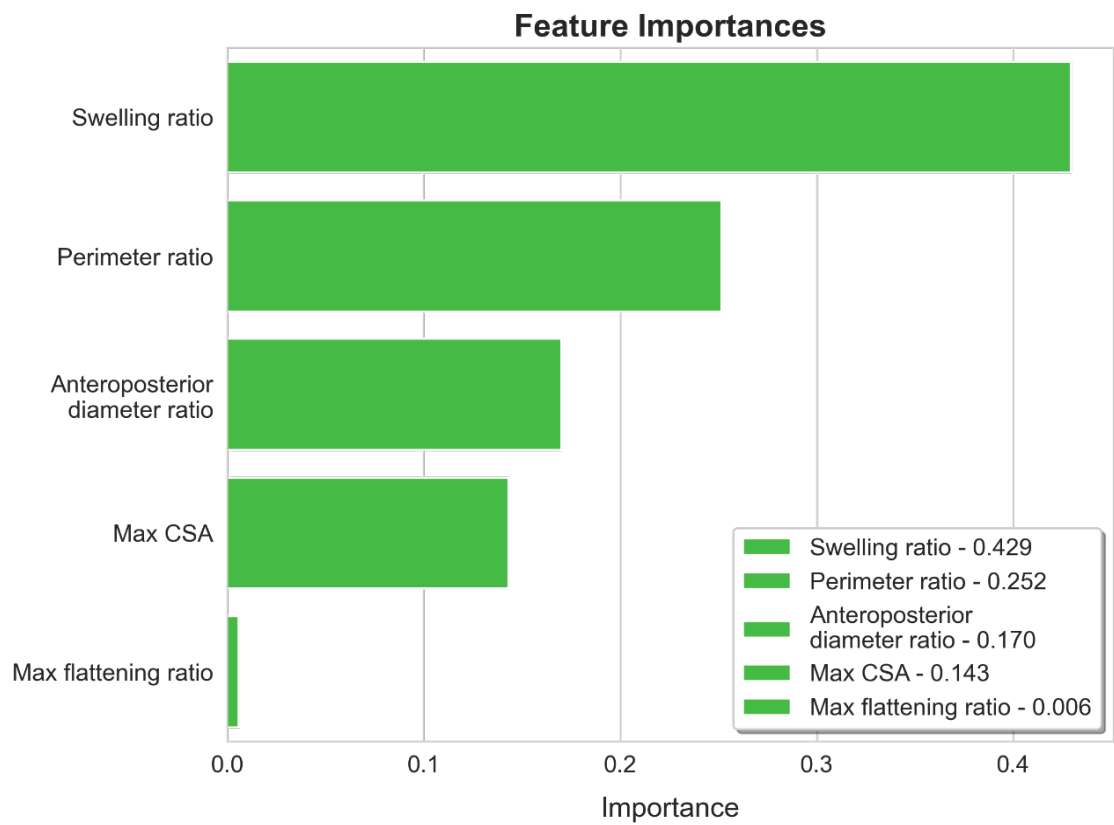

Figure 7. Feature importances of diagnostic parameters in OSA-CTSD.

Furthermore, we analyzed the contribution ratios for each indicator in the OSA-CTSD method (Fig. 7). As shown in Figure 7, it showed a strong correlation between the feature contribution of a single indicator and its diagnostic accuracy when used alone (Table 5). And the contribution value for SR was much higher than that of other indicators with a percentage as high as 42.9%. This also confirms that CTS pathogenesis involves compression on the MN within the carpal tunnel leading to secondary edema near proximal or/ and distal ends. However, FR performed poorly, possibly because when the MN is significantly compressed in a narrow segment, both its transverse diameter and longitudinal axis are often simultaneously compressed. According to literature reports, SR had high specificity for diagnosing CTS(8, 9, 33).

However, due to significantly increased workload and low reproducibility among US radiologists, this method has not been used frequently. On the contrary, the proposed OSA-CTSD can automatically generate accurate and reproducible SR measurements without manual interventions. Furthermore, this study is currently unique among AI-assisted US diagnoses for CTS by providing comprehensive analysis across the proximal entrance, carpal tunnel segment, and distal exit segments, which can better fit the CTS pathogenesis mechanism while reducing the misdiagnosis rate for US diagnosis of CTS.

**Conclusions**

In this study, we proposed a one-stop automated CTS diagnosis system (OSA-CTSD) as an effective CAD tool for diagnosing CTS based on US. The OSA-CTSD combined three processes into a unified framework, including real-time MN delineation, accurate biometric measurements, and explainable CTS diagnosis. The proposed tool is evaluated on a large-scale dataset including 32,301 images from 90 normal wrists and 40 CTS wrists, and by multiple metrics. It demonstrated promising diagnostic performance based on clinical-interpretable parameters. Besides, it is fully automated with a simplified scanning protocol. The application of such a tool could not only reduce reliance on the expertise of examiners, but also could help to promote the standardization of the CTS diagnosis process in the future.

Acknowledgments—This work was supported by the National Natural Science Foundation of China (No. 62101342, and No. 62171290); Guangdong Basic and Applied Basic Research Foundation (No.2023A1515012960); Shenzhen-Hong Kong Joint Research Program (No. SGDX20201103095613036) and Shenzhen Key Medical Discipline Construction Fund (SZXK052). Authors Jiayu Peng, Jiajun Zeng, and Manlin Lai contributed equally to this work.

Conflict of interest—None of the authors has a financial conflict related to the content of this work.